\documentclass[12pt]{article}
\usepackage{graphicx}
\usepackage{ae,aecompl} 
\usepackage{amsmath,amsfonts,amssymb}
\usepackage{authblk}
\usepackage{cite}
\date{}
        
\begin{document}
\title{Entropy facilitated active transport}
\author[$\dagger$, $\ddag$]{J. M. Rub{\'i}}
\author[$\ddag$]{A.  Lervik\footnote{anders.lervik@ntnu.no}}
\author[$\ddag$]{D. Bedeaux}
\author[$\ddag$]{S. Kjelstrup}
\affil[$\dagger$]{Statistical and Interdisciplinary Physics Section,
Department de F{\'i}sica de la Mat{\`e}ria Condensada,
Universitat de Barcelona, Mart{\'i} i Franqu{\`e}s 1, 08028,
Barcelona, Spain}
\affil[$\ddag$]{Department of Chemistry, Faculty of Natural Sciences and
Technology, Norwegian University of Science and
Technology, Trondheim, Norway}
\maketitle
\begin{abstract}
{We show how active transport of ions can be interpreted as
an entropy facilitated process. In this interpretation, the
pore geometry 
through which substrates are transported can give rise
to a driving force. This gives
a direct link between
the geometry 
and the changes in Gibbs energy required. Quantifying the size of
this effect for several proteins we find that the entropic contribution
from the pore geometry is significant and we discuss 
how the effect can be used to interpret variations in the affinity at the binding site.}
\end{abstract}

\noindent Active transport is of major importance
in biology; meaning transport of a compound
against its chemical potential, driven by a chemical
reaction. A notable example is the large P-type ATPase
protein family which functions as ion or lipid pumps, 
crucial for a wide range of processes in almost all
forms of life~\cite{palmgren2011}.
The P-type ATPases share a common topology and their operation
can be described using a Post-Albers cycle
with four key conformations: 
E1, E1P, EP2 and E2~\cite{Sitsel2015}.
The transitions E1P$\leftrightarrow$E2P and E2$\leftrightarrow$E1 in the catalytic
cycle are associated with large conformational changes~\cite{Sitsel2015}
and it is clear that the energy released by 
ATP hydrolysis drives the enzyme between these states.

One may ask about the meaning of the conformational changes.
In particular, the formation of a wide funnel-shaped 
outlet channel can be observed in several
P-type ATPases~\cite{olesen2007, takeuchi2009,wang2014, Abe1637}
when the enzyme changes from the E1P state to the E2P state. 
The occluded ion is then eventually
exposed to the lumen; but why does the ion leave
its binding site when the chemical potential is so
much larger in the direction of transport than in the 
opposite direction?
Can this be influenced by the observed pore geometries?
How can the change in Gibbs energy
by the chemical reaction, which may take
place relatively far away from the
binding site (see e.g. Ref.~\cite{Inesi2011}),
be transferred and used at the ion binding site? These
questions, which have
been of major interest since the discovery of the pumps, are
still not fully answered
and this is the topic we discuss here.

We shall argue that active transport 
can be better understood by
shifting focus from an energy to an entropy barrier in
a specific part of the translocation process.
When the outlet channel is formed in the E2P state,
the shape of this channel directly facilitates
the transport of ions by allowing for an entropy
increase, or equivalently, a 
chemical potential decrease in the E2P state as we will
demonstrate.
The idea that the actual pore geometry
may play a role in active transport,
stems from previous studies which have shown how entropic 
barriers can play a major
role for separation purposes~\cite{reguera2001, rubi2010, reguera2012}.

\begin{figure}[tbp]
\centering
\includegraphics[width=0.65\textwidth]{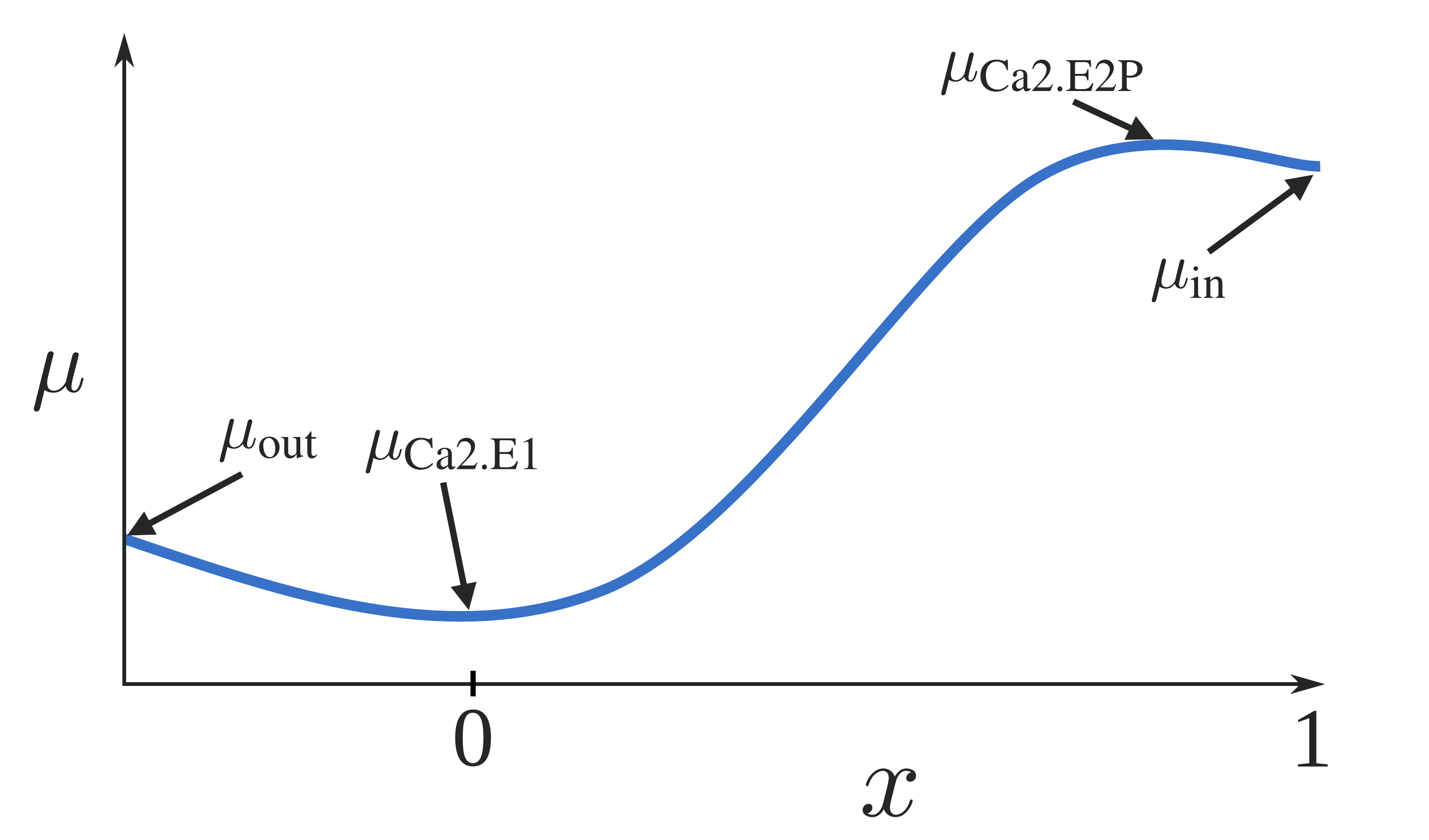}
\caption{(Color on-line.) 
Illustration of the variation in the chemical potential $\mu$ of a
Ca$^{2+}$-ion as it is
transported by the Ca$^{2+}$-ATPase from
the cytosol to the lumen of SR.
The coordinate $x$
denotes the extent of the transport inside funnel-shaped
pore from the binding site at $x=0$ to the lumen
at $x=1$.
The chemical potential is indicated for four states;
in the cytosol/outside ($\mu_\text{out}$), 
lumen/inside ($\mu_\text{in}$), the E1 ($\mu_\text{Ca2.E1}$) and the E2P ($\mu_\text{Ca2.E2P}$) states.
For a concentration ratio of $10^4$ at 
$37^\circ$C, the energy requirement is
$24$~kJ per mol ion transported.
}
\label{fig:chempot2}
\end{figure}

To exemplify the impact of the pore geometry, we will
take the Ca$^{2+}$ transporting Ca$^{2+}$-ATPase
of sarcoplasmic reticulum (SR) but we note that the ideas
presented may apply more generally. 
For the Ca$^{2+}$-ATPase, we illustrate the variation of the
chemical potential
as ions are transported from the cytosol to the lumen of the SR 
in Fig.~\ref{fig:chempot2}.
It is known that the binding of Ca$^{2+}$ is fast~\cite{peinelt2005}
and at equilibrium, the chemical potential of Ca$^{2+}$ in the E1 state,
$\mu_{\text{Ca2.E1}}$, is 
equal to the chemical potential in the cytosol, $\mu_{\text{out}}$.
During operation of the pump, there is
probably a small difference between $\mu_{\text{out}}$ and 
$\mu_{\text{Ca2.E1}}$ but the chemical potential in the final
state, $\mu_\text{in}$, is clearly larger than $\mu_{\text{out}}$
as depicted in Fig.~\ref{fig:chempot2}.
There is a large uncertainty 
related to the chemical potential of Ca$^{2+}$ 
when the enzyme is in the state E2P,  $\mu_{\text{Ca2.E2P}}$.
In Fig~\ref{fig:chempot2} it is assumed to be close to 
$\mu_\text{in}$ (which is the lower bound resulting in a positive
Ca$^{2+}$ flux), enabling the ion to pass to the lumen.
The variation in binding energy inherent in this picture
has been attributed to the
enthalpic part of the chemical potential, as 
the enthalpy gives a measure of the bond strength:
Repulsive forces 
can raise the chemical potential 
$\mu_\text{Ca2.E2P}$, which results in a low affinity binding site.
In such a situation the ions can move spontaneously to the lumen.
Alternatively, this change in chemical potential can be attributed
to entropic effects.
The leap from $\mu_\text{Ca2.E1}$ to $\mu_\text{Ca2.E2P}$ is
due to the hydrolysis of ATP
and it can arise from increasing the enthalpy of
the ion, or by lowering
the entropy of the ion.
The latter can be brought about by
a change in the actual shape
of the channel where the ion is transported.
A particle moving in a pore with a varying cross-sectional area
will be subjected to entropic forces due to the change
in area and this can in fact \emph{facilitate} the
transport. 

\begin{figure}[tbp]
\centering
\includegraphics[width=0.65\textwidth]{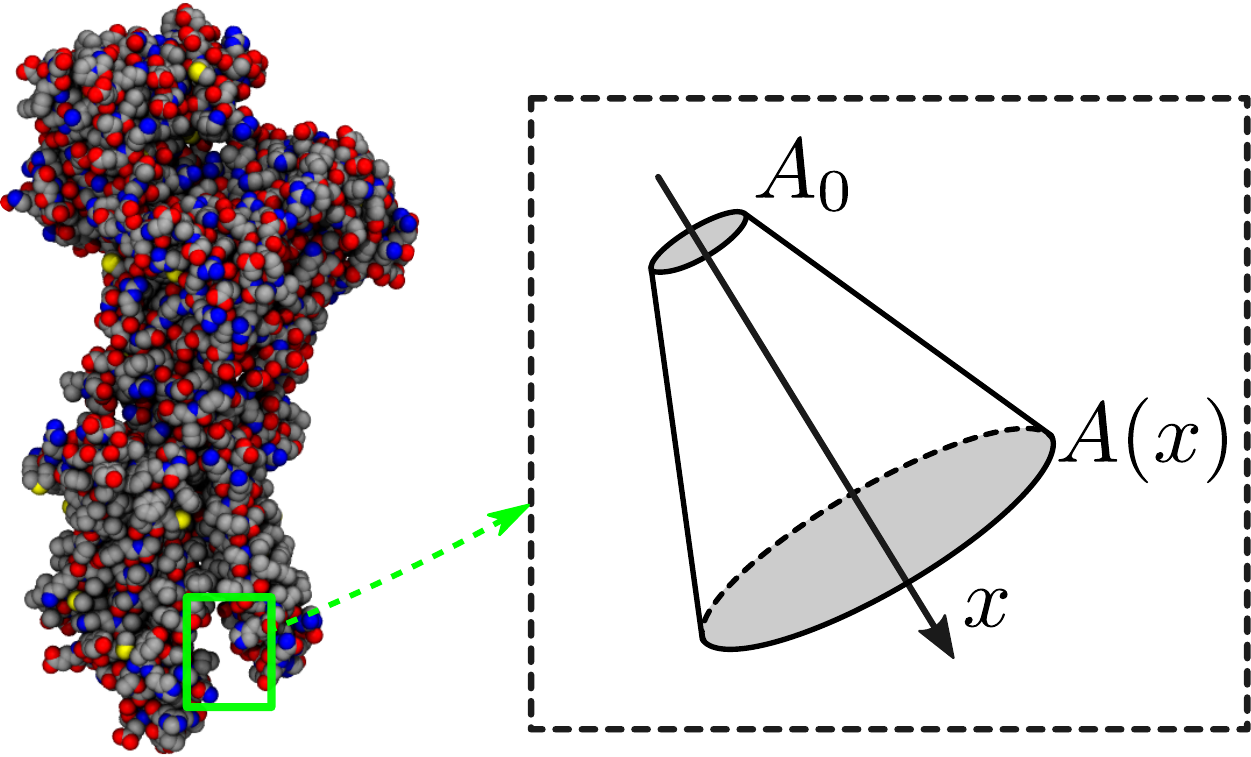}
\caption{(Color on-line.) 
The E2P state for the Ca$^{2+}$-ATPase
(Protein Data Bank ID: 3B9B~\cite{olesen2007}) showing
the funnel shaped channel. The inset shows a conical funnel where the
area increases from the start of the channel ($A_0$) to the exit ($A_\text{in}$) where
$0\leq x \leq 1$ indicates the position along the channel.}
\label{fig:channels}
\end{figure}

The outlet channels observed in the P-type ATPases
are narrow but widens towards the exit of the channel,
as illustrated in Fig~\ref{fig:channels}. We assume that the
ion transport is effectively one-dimensional, 
directed along a spatial coordinate $x$.
The ion density along the transport direction may
be approximated by~\cite{reguera2001}
\begin{equation}
\rho(x) = \rho_{0}\frac{A(x)}{A_{0}}
\label{eq:density0}
\end{equation}
where $\rho_{0}$ and $A_{0}$ are the ion density and cross-sectional area at a reference
point, say $x=0$.
The size of the particles has been shown to influence the
transport properties~\cite{riefler2010}
and here, both $A(x)$ and $A_0$ measures the available area, e.g. for a
circular geometry: $A(x) = \pi \left(r(x) - a \right)^2$ where $r(x)$ is the
radius of the pore at position $x$ and $a$ the radius of the
ion.
The density given by Eq.~\ref{eq:density0} can also be viewed
as the canonical distribution function
\begin{equation}
\rho(x) = \rho_{0} \exp \left( \frac{\Delta S_{\text{pore}} (x) }{R}\right)
\end{equation}
where we have introduced the change in entropy, $\Delta S_{\text{pore}} (x)$,
associated with the change in cross-sectional area.
From these equations, we can conclude that change in the entropy 
during ion motion along the pore is directly related to the
cross sectional area
\begin{equation}
\Delta S_{\text{pore}}=R \ln \frac{A(x)}{A_{0}}
\label{eq:A}
\end{equation}
and that the position dependent entropy along the ion trajectory
is giving rise to a (thermodynamic) force of entropic
nature, $F_{\text{ent}}$, acting on the ions
\begin{equation}
F_{\text{ent}}= T 
\frac{\partial \Delta S_{\text{pore}}}{\partial x}=
RT\frac{1}{A(x)}\frac{\partial A(x)}{\partial x}
\end{equation}
The direction of the force directly depends on the slope of the channel.
For the conical structure shown in Fig.~\ref{fig:channels} the entropic force is positive, 
and will facilitate translocation in direction of the wider opening.
The overall entropy change ($\Delta S$) for the ion translocation will thus have
two contributions: the normal contribution from the ion
concentration difference
($\Delta S_{\text{Ca}}=-R\ln(c_\text{in}/c_\text{out}) = - \Delta \mu_{\text{Ca}}/T$),
and a special contribution from the pore
shape. Per mol of ion we have $\Delta S = \Delta S_{\text{Ca}} + \Delta S_{\text{pore}}$
and by
introducing Eq.~\ref{eq:A} for $x=1$, we obtain
\begin{equation}
\Delta S = R  \ln \frac{c_{\text{out}}A_{\text{in}}}{c_{\text{in}}A_{\text{0}}}
\label{eq:dgpore0}
\end{equation}
The entropic force depends on the concentration ratio,  
but also on the ratio of the cross-sectional areas at the
two sides of the channel. 
A large $c_{\text{in}}/c_{\text{out}}$ ratio may be counteracted by a
smaller $A_0/A_{\text{in}}$ ratio, meaning that translocation may take place against the
gradient in concentration. 
In other words, it may be facilitated by entropic forces induced 
by the formation of a conical pore and the larger the ratio 
$A_{\text{in}}/A_0$, the larger is the co-acting  entropic force.

The translocation rate, $J_\text{Ca}$, 
can be shown to be
\begin{equation}
J_\text{Ca} =
- L \left[ 1 - 
\left(\frac{c_{\text{out}}A_{\text{in}}}{c_{\text{in}}A_{\text{0}}}\right)
\exp \left( -\frac{\Delta H}{RT}\right)
\right]  
\label{eq:fluxpore0}
\end{equation}
where $L$ is the backward reaction rate,
and $\Delta H$ is the change in enthalpy.
The equation holds when the activity coefficients inside and outside the membrane are the same.
In the outset $\Delta H$ is unknown.  In the case that the pump operates far
from equilibrium the last term in the parenthesis in Eq.~\ref{eq:fluxpore0} dominates, giving
  \begin{equation}
J_\text{Ca} =  L 
\left(
\frac{c_{\text{out}}A_{\text{in}}}
{c_{\text{in}}A_{\text{0}}}
\right)
\exp\left( -\frac{\Delta H}{RT}\right) 
\label{eq:fluxpore2}
\end{equation} 
The equation gives an Arrhenius behavior of
the current through the dependence on $\Delta H$. 
Assuming that the entropy change alone drives the translocation, we have
\begin{equation}
J_\text{Ca} =  L \left( \frac{c_{\text{out}} A_{\text{in}}}{c_{\text{in}}A_{0}}\right)
\end{equation}
where the flux is given by the ratio of concentrations and cross-sectional areas.

\begin{figure}[!tbp]
\centering
\includegraphics[width=0.65\textwidth]{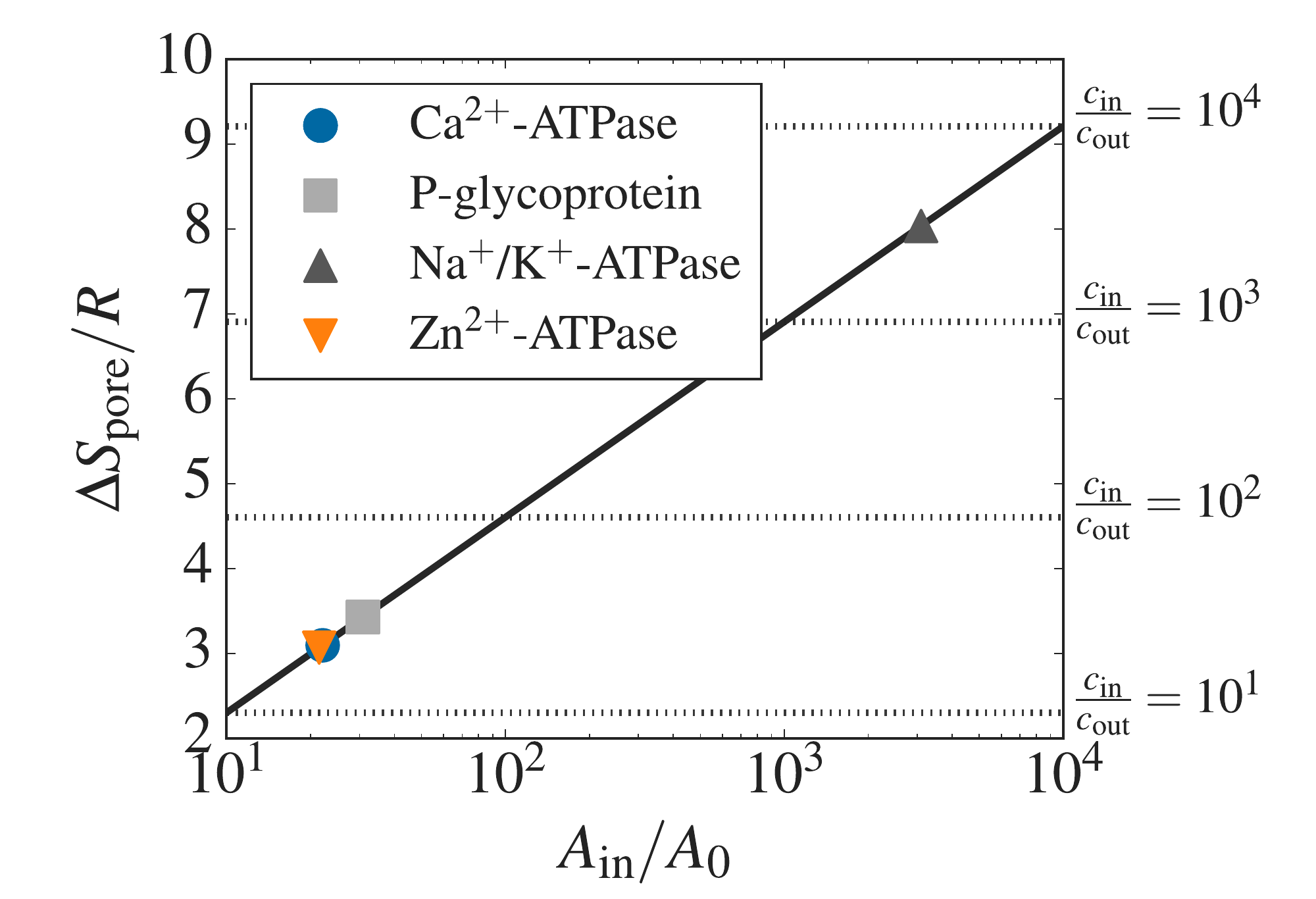}
\caption{(Color on-line.)
Entropic contribution, $\Delta S_{\text{pore}}$,
(solid black line) to the driving force
as a function of the ratio of cross sectional areas.
Contributions for several enzymes
are also shown:
Zn$^{2+}$-ATPase~\cite{Sitsel2015}
P-glycoprotein~\cite{Loo2001},
Ca$^{2+}$-ATPase~\cite{olesen2007} and
Na$^+$/K$^+$-ATPase~\cite{takeuchi2009}.
For the Zn$^{2+}$-ATPase, the widths of the channel were estimated
using the crystal structure (Protein Data Bank ID: 4UMV).
We have considered the size of the ions using the ionic radii
reported by Marcus~\cite{marcus1988}
and for the Na$^+$/K$^+$-ATPase we only considered the smallest ion (Na$^+$).
As P-glycoprotein may transport
many different compounds, we have omitted the size of the
transported compound and the estimate represents a lower bound.
}
\label{fig:results}
\end{figure}

Assuming a circular cross-sectional area,
the entropic driving force is,
\begin{equation}
- T \Delta S = R T \ln \left[ \left( \frac{c_{\text{in}}}{c_{\text{out}}} \right)
\left(\frac{r_{0}-a}{r_{\text{in}}-a}\right)^2
\right]
\end{equation}
where $r_{0}$ is the radius at the start of
the channel, $r_{\text{in}}$ at the exit of the channel and
$a$ the radius of the ion. 
In Fig.~\ref{fig:results} we show the entropic contribution to the
 driving force for several enzymes. As the results show, 
the entropic contribution can be sizable. It is not sufficient to explain
active transport by itself for all the enzymes as can be
seen by the results for the Ca$^{2+}$-ATPase. In this case, 
the entropic contribution is on the order of $10$~kJ/mol ion transported at $37^\circ\text{C}$.
This accounts for $40\%$ of the chemical
potential difference which indicate that other contributions, for instance enthalpic effects,
may also be important for this protein.
For the Na$^+$/K$^+$-ATPase we find that, 
due to the narrowness of the
pore, the entropic contribution ($20$~kJ/mol) is relatively
large compared to the other enzymes.
This is sufficient to overcome the concentration
ratio of Na$^+$ at physiological conditions, however,
in this case, the Na$^+$ ions are transported against
the resting membrane potential.
Assuming a resting potential of $-60$~mV and
a concentration ratio on the order of $10$, the required Gibbs energy
is $12$~kJ/mol ion transported. This means that the entropic force
estimated in Fig.~\ref{fig:results} is still sufficient for this enzyme.
The entropic contribution for the two other enzymes is
similar to the Ca$^{2+}$-ATPase. We have neglected the possibility
of complexing the ions with other species such as water which could
increase the effective radius and the entropic force.

We have shown that the actual pore geometry observed in several transport
enzymes may give rise to an entropic force, facilitating transport.
This entropic effect can be used, together with enthalpic effects, to explain how
conformational changes influences the binding site, allowing transport of
compounds against their concentration gradients. In particular, the entropic force
gives a direct link between the energy released by a reaction at the active site
and changes in the chemical potential at the binding site: By modifying the conformation
and the pore geometry, a low affinity binding site can be created.

\section*{Author contributions}
JMR, AL, DB, and SK developed the theory,
performed the analysis, and wrote the manuscript. 
JMR initiated the work.

\section*{Acknowledgements}
The Norwegian University of Science and Technology
is thanked for supporting the stay of JMR.

\end{document}